\documentstyle[psfig]{l-aa}
\begin{document}
\thesaurus{  }
\title{\bf Large bent jets in the inner region of CSSs}   
\author{F. Mantovani\inst{1}, W. Junor\inst{2}, M. Bondi\inst{1}, 
W. Cotton\inst{3}, R. Fanti\inst{1,4}, L. Padrielli\inst{1}, 
G.D. Nicolson\inst{5} \& E. Salerno\inst{1}}
\institute{ Istituto di Radioastronomia del C.N.R., Via P. Gobetti, I-40129
\and Institute for Astrophysics, University of New Mexico, NM, USA
\and National Radio Astronomy Observatory, Charlottesville, USA
\and Dip. di Fisica, Universit\'a degli Studi, Bologna, Italy
\and Hartebeesthoek Radio Astronomy Observatory, Krugersdorp, South Africa
}
\offprints{Franco Mantovani - fmantovani@ira.bo.cnr.it}
\date{}
\maketitle
\markboth{Mantovani et al.}{Large bent jets in the inner region of CSSs}   
\begin{abstract}
The class of Compact Steep-spectrum Sources is dominated by 
double-lobed objects (70\%). The remaining 30\% are jet-dominated objects,
with the jet
brightened either by Doppler boosting or by interaction with the ambient
media.  We show that there is both observational and statistical evidence
in favour of an interaction between jets and dense gas clouds.
Such an interaction should happen in the Narrow Line Regions.
The images of four CSSs observed by us with VLBI are also presented.
These sources do show large bent jets in the first kpc from the
nucleus.
\end{abstract}
\section{Introduction}

The class of Compact Steep-spectrum Sources (CSSs) has been discussed by
many authors (e.g.\ Pearson et al.\ 1985; Fanti it et al.\
1990).  It is now generally believed that CSSs are physically small
objects with sub-galactic dimensions, whose structure and sizes are
possibly affected by the ambient gas in the central regions of the
parent optical objects.  Double-lobed sources represent $\sim 70\%$ of
the objects in a sample of CSSs (Fanti et al.\ 1995)
selected from the 3CR catalogue (Laing et al.\ 1983) and from the 
Peacock \& Wall (1982) catalogue for a linear size $<15$\,kpc,
spectral index $\alpha>0.5$ (S$\propto\nu^{-\alpha}$) and radio power 
$\geq10^{27}$\,W\,Hz$^{-1}$ at 178\,MHz for the 3CR or
$\geq10^{26}$\,W\,Hz$^{-1}$ at 2.7\,GHz ($H_{0} = 100
$km\,sec$^{-1}$Mpc$^{-1}$ and $q_{0} = 1$).   
Fanti at al.\ (1995) suggest that double-lobed CSSs represent the {\it
young}  ($\leq10^6$ yrs) precursors of the larger double-lobed radio
sources. The remaining  30\% might represent a different
population, made of sources where the jets are brightened either by Doppler 
boosting or by interaction with the ambient medium. 

There are observational results that suggest a connection between
radio source evolution and the CSS and GPS (GigaHertzed Peaked
Sources) phenomena (see, for example, O'Dea et al.\ 1991; Stanghellini et
al.\ 1992). These results also support the hypothesis that the CSS
phenomenon results from expansion within the ambient environment of
radio sources at intermediate and high redshifts.  This may be due to
the existence of a more dense ISM in the inner portions of host
galaxies with CSSs.  This density enhancement would influence the
behaviour of the radio-emitting regions. 

Generally speaking, the cores are often weak in CSSs, suggesting that
if boosting is present, then jets and cores are misaligned with  
respect to the kpc-scale jets (Spencer 1994).  Moreover, distorted
structures on the milli-arcsecond scale are common in those CSSs which
are quasars with strong jets. 

In this paper we focus our attention on a sample of CSSs selected because
of the large bends in the jets in their inner regions.   We will
consider whether the large misalignments are due to relativistic
effects or produced by jet-cloud interactions in Narrow Line
Regions. In the next section we will describe the sample, our
observations of four of the selected sources and the source structures.
We will then summarize the source parameters derived from the
observations, and estimate the probability of the collision of a jet
with dense gas cloud.  Finally, we will discuss the results of our
investigations.

\section{The sample}

We have used the images of sources classified as CSSs from the
compilation by Dallacasa \& Stanghellini (1990).  From this, we have
chosen a sub-sample using quite simple criteria.  First, all the
sources dominated by jet emission were selected. From this, a  
list of sources was chosen to include all the sources showing large
bent jets on the sub-arcsecond scale.  A bend in a jet is
considered {\it large} when the jet major axis changes direction 
by an angle $>50^{\circ}$ ($\Delta PA >50^{\circ}$).

Such a value was adopted to avoid ambiguities in selecting candidates
from the published images.  These are due to the uncertainty in
determining the Position Angle (PA) of the jet major axis.  We are
also aware that the images were made from observations done using a
variety of arrays and frequencies, some of which were inappropriate
for our purpose.  The selected sources are listed in Table~3
below. Observations of four of them by us are presented here.

\subsection{The observations}

The four sources (0548$+$165, 1741$+$279, 2033$+$187 and 2147$+$145)
have been observed with VLBI at 1.6\,GHz and 5\,GHz using different
arrays and recording systems.  Calibration observations on
largely-unresolved radio sources were made during each observing
session along with the observations of the target source.
Table~1 summarizes the observations.  After the correlation
process the bulk of the data reduction was done using the {\em AIPS} 
package.  The sources have been imaged using both {\em AIPS} and {\em
DIFMAP} (Shepherd et al.\ 1995). 

The source parameters in Table~2 are as follows: 
-- column 1: source name; column 2: observing frequency;
columns 3,4,5: beam major axis, minor axis in mas and PA in degrees;
column 6: r.m.s.\ noise in the map far from the source; column 7:
component label; column 8,9,10: major axis, minor axis in mas and PA
in deg of each component; column 11: component peak brightness in
mJy/beam; column 12: component total flux density in mJy; column 12:
component spectral index. 
%
%
%
\begin{table*} 
\small
\begin{center}
\caption{Observing information}
\medskip
\begin{tabular}{|l|l|r|l|l|l|l|} 
\hline  
Source     & Array         & Freq. & Term. & Obs.Date  & Track.&  Corr.   \cr
\hline 
0548$+$165 & B,S,J2,L,I,O,W &  5 GHz & MK3-B & 12May1995 & 12 hrs & MPIfR   \cr
           & B,S,J1,L,O,W,K,H,G,  &1.6 GHz & MK2   & 21Sep1992 & 04 hrs& CIT\cr
           & VLBA                 &        &       &           &       &    \cr
1741$+$279 & B,J2,L,I,O,W &  5 GHz & MK3-B & 14Sep1993 & 10 hrs & MPIfR     \cr
           & B,J1,L,O,W,K,H,G,  &1.6 GHz & MK2   & 21Sep1992 & 09 hrs  & CIT\cr
           & VLBA               &        &       &           &       &    \cr
2033$+$187 & B,J2,L,I,O,W &  5 GHz & MK3-B & 01Mar1994 & 10 hrs & MPIfR     \cr
           & S,O,W,C,D,K,H,G,  &1.6 GHz & MK2   & 14Mar1986 & 08 hrs   & CIT\cr
           & Hr,Fd,Ov          &        &       &           &          &    \cr
2147$+$145 & B,L,O,W      &  5 GHz & MK3-B & 26Feb1994 & 11 hrs & MPIfR     \cr
\hline
\end{tabular}
\end{center}
%
{\em Note:}~~{\bf B} Effelsberg, {\bf C} Cambridge, {\bf D} Defford, 
             {\bf Fd} Fort Davis,
             {\bf G} Green Bank, {\bf H} Hartebeestehoek, {\bf K} Haystack,
             {\bf Hr} Hat Creek,
             {\bf J1} Jodrell-Lovell, {\bf J2} Jodrell-MK2, 
             {\bf L} Medicina, {\bf I} Noto, {\bf O} Onsala85, 
             {\bf Ov} Owens Valley,
             {\bf S} Simeiz, {\bf W} Westerbork, 
             {\bf VLBA:} PT,KP,LA,BR,FD,NL,OV 
\end{table*}
%
%
%
%
%
\begin{table*} 
\small
\begin{center}
\caption{Sources parameters}
\medskip
\begin{tabular}{|lrrrrrlrrrrrr|} 
\hline  
Source& Obs.&         &Beam     &      & rms    & C &    &De.Size&  & Flux 
&Dens.  &$\alpha$ \cr
      &$\nu$& maj     &   min   & PA   & noise  &   &maj &min & PA & peak &
total &   \cr
      & MHz & mas     & mas     &$\deg$& mJy/b  &   &mas&mas&$\deg$&mJy/b& mJy   &        \cr
\hline 
0548$+$165& 1662& 14& 10 & 8 & 10 & a &- & -& -& 116   & 131   & 0.2    \cr
          &     &   &    &   &    & j & -& -& -&  62   & 162   &        \cr
          &     &   &    &   &    & b &20& 9&15& 693   &1571   & 1.3    \cr
          &     &   &    &   &    & c & -& -& -&  16   &  18   &        \cr
          & 4975& 6 & 6  &   & 0.2& a & -& -& -&  92.6 & 109.1 &        \cr
          &     &   &    &   &    & j & -& -& -&   8.3 &  49.1 &        \cr
          &     &   &    &   &    & b &14& 9& 9& 110.9 & 376.5 &        \cr
          &     &   &    &   &    & c &21& 9&34&  15.8 &  92.0 &        \cr
1741$+$279& 1662& 12& 9  &--2&2.5 & a & -& -& -&  53   &  57   & $-$0.3   \cr
          &     &   &    &   &    & b &29&11&104& 10   &  44   &  1.3   \cr
          &     &   &    &   &    & c &36& 9&110&  6   &  25   &  1.8   \cr
          &     &   &    &   &    & d & -& -& -&  18   &  23   &  1.3   \cr
          &     &   &    &   &    & e &13& 8& 7&  77   & 178   &  1.4   \cr
          & 4960&8.5& 5.7& 79&0.12& a & 4& 2&69&  75.9 &  82.9 &      \cr
          &     &   &    &   &    & b &13&12&108&   3.3 &  10.7 &   \cr
          &     &   &    &   &    & c & 6& 2&172&   2.3 &   3.5 &     \cr
          &     &   &    &   &    & d &16&11&72&  18.5 &  42.7 &    \cr
          &     &   &    &   &    & e &12& 6&29&  18.5 &  42.7 &     \cr
2033$+$187& 1662& 19& 3  &--9& 0.7& a &8 &4 &164& 303.2 & 499.2 &  1.2\cr
          &     &   &    &   &  &a$_1$&33&3 &164&   12.6& 23.9  &    \cr
          &     &   &    &   &  &a$_2$&12& 4&33&   64.0&154.8  &     \cr
          &     &   &    &   &    & b &10& 3&175&  63.5 &133.9  & 1.1\cr
          &     &   &    &   &    & c &--&--&--&  13.6 & 14.4  &        \cr
          &     &   &    &   &    & d &10 &5 &168&  80.3 &193.2  & 1.2\cr
          & 4960&9.7&4.8 &65 &0.07& a &7& 5&71&  94.8 &136.0  &      \cr
          &     &   &    &   &    & b &7& 1&79&  33.1 & 38.6  &      \cr
          &     &   &    &   &    & d &7.3&4.4&68&  10.6 & 12.6  &      \cr
          &     &   &    &   &  &d$_1$&--&--&--&   29.9& 41.5  &        \cr
2147$+$145& 4960&12.6&8.2&70 & 0.2& a &--&--&--& 394.5 &486.5  &        \cr
          &     &    &   &   &    & b &19& 6&55&  40.5 & 56.5  &   \cr
          &     &    &   &   &    & c &14& 7&107&   9.1 &  9.6  &    \cr
          &     &    &   &   &    & d &17&1 &153&  23.8 & 61.5  &    \cr
          &     &    &   &   &    & e &--&--&--&   5.0 &  3.6  &        \cr
          &     &    &   &   &    & f &--&--&--&   6.7 &  6.1  &        \cr
\hline
\end{tabular}
\end{center}
%
%
\end{table*}
\subsection{Description of Sources}

The four sources above plus the other sources listed in the following 
Tab.\,3 are described briefly here:

{\bf 3C43} (0127$+$233)

\noindent
The VLA image from Pearson et al.\ (1985) shows a highly misaligned
triple structure. The MERLIN map by Sanghera et al.\ (1995) shows that
the northernmost component A in the EVN $\lambda$18cm image by Spencer
et al.\ (1991) is likely to be the core. This is consistent with the
$\lambda$50cm VLBI observations of Rendong et al.\ (1991). Spencer
(1994) imaged 3C43 at $\lambda$18cm with MERLIN.  That image shows a
bridge of emission between the main components and the northern
component. The jet is straight and collimated for the first 150 mas,
then it changes PA by $33^{\circ}$ and, after 75 mas the PA changes
again by $60^{\circ}$ pointing towards east.  According to Junor et
al.\ (in preparation) the central component is 3.6\% polarized
at 8.4\,GHz with the VLA. The polarized intensity is sufficiently
strong to provide an estimate of Rotation Measure, RM, of
$-1800\,$rad~m$^{-2}$ in the source's rest frame. The intrinsic
magnetic field direction in the central component follows the
curvature of the source faithfully.

{\bf 3C99} (0358$+$004) 

\noindent
The radio source 3C99 has a triple structure on arcsecond scales. The
outer components are located rather asymmetrically relative to the
nucleus and have very different surface brightnesses (Mantovani et
al.\ 1990). Along the major axis, 3C99 has an angular size of
$\simeq$6 arcsec and a linear size of $\simeq$21 kpc. It is associated
with an N~galaxy which has been detected close to the central component
(Spinrad et al.\ 1985). The central component is unpolarized in
8.4\,GHz VLA observations (Mantovani et al.\ 1997) and it is likely 
to be the nucleus of 3C99.  The VLBI image of the source at
$\lambda$18cm shows that the central component consists of several
blobs of emission with the two prominent ones being significantly
misaligned with the overall axis of the source.  

{\bf 3C119} (0429$+$415)

\noindent
An image of the source structure with an angular resolution of 5 mas is 
presented by Nan Ren-dong et al.\ (1991). Component A has an inverted
spectral index and it is identified with the core of 3C119.  The
morphology of 3C119 is rather peculiar.  The jet emerging from the
core component is not well collimated. It contains several blobs of
emission. The major axis PA changes direction by $55^{\circ}$ to reach
the component C, at about 40 mas from the core. From there, it changes
direction several times to form an almost circular structure. The core
has a radio luminosity which is $< 2\%$ of the source total luminosity
at 5 GHz.  Taylor et al.\ (1992) listed 3C119 among sources with very
large Rotation Measures (RM$=$3400~rad~m$^{-2}$). A number of possible
explanations have been discussed in the paper by Nan Ren-dong et al.\
(1991) for the source's brightness distribution (e.g.\ rotational
shear of the radio jet by an ambient rotating gaseous disk, precession
in the nucleus, and the source expanding in a cavity in the
interstellar medium), but those authors admit that they cannot reach a
satisfactory conclusion. 

{\bf 3C147} (0538$+$498)

\noindent
3C147 has been observed by several investigators; see, for example,
the collection of images by Alef et al.\ (1990).  Those authors have
also observed 3C147 with 5\,GHz VLBI over three epochs. The source 
shows an unusually-complex, nonlinear structure which varies with
time.  Superluminal separation of two components in the core region
was observed also.  New 8.4\,GHz data (Alef, private communication)
confirm a mildly-superluminal separation velocity of $v_{app}\sim1.3
c/h$.   The jet is embedded in a diffuse emission region and  shows a
change in the projected orientation of its major axis of $90^{\circ}$
at $\sim200$ mas from the core. A VLA image at 1\, GHz  (van Bruegel et
al.\ 1984) shows a weak component north to the main one in a position
which is opposite to the jet respect to the core.  3C147 shows large
RM's of $-3144$ and $+630$ rad m$^{-2}$ in the rest frame of the
source for the main component and for the extension to
the NNE respectively (Junor et al.\ submitted).

{\bf 0548$+$165}

\noindent
On arcsecond scales, the source shows an asymmetric structure with a
strong, unresolved component coincident with a quasar at z=0.474 and a
much weaker secondary component about 3 arcsec to the north. The
faint, extended component is weakly polarized.  Most of the
polarization comes from the main  component. This source has
a RM of 1934 rad m$^{-2}$. The polarized emission is also strongly
depolarized between 15 GHz and 5 GHz (Mantovani et al.\ 1994). 

The mas scale structure of 0548$+$165 from a full-track, global
$\lambda$ 18cm program is shown in Fig.\,1. The source has a straight
jet pointing west which changes direction dramatically $90^{\circ}$ at
$\sim$80 mas from the core. 
\begin{figure}
\psfig{figure=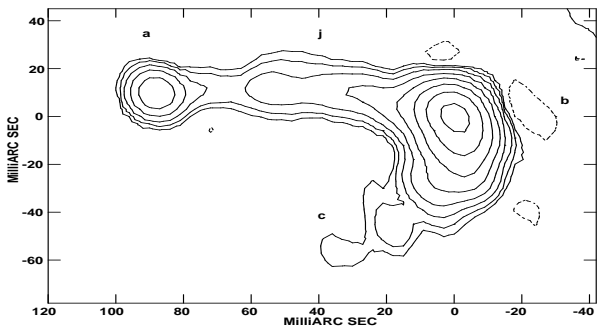,height=8.9cm,width=8.9cm}
\caption[]{ VLBI image of 0548$+$165 at 18\,cm. Contours are -4,
4, 8, 16, 32, 64, 128, 256, 512\,mJy/beam. The peak
flux density is 692.6\,mJy/beam.}
\end{figure}

The de-rotated magnetic field is aligned parallel to the east-west
direction (Mantovani et al.\ 1994) like the straight part at the
beginning of the jet.  

EVN observations at $\lambda$ 6cm (Fig.\,2) confirm the main structure 
seen at $\lambda$18cm.  Fig.\,2 also shows that the jet increases
its width but it remains collimated and it seems to show a wiggling
structure. The core is thought to be the easternmost component, since
it has a flat spectrum. The detection of the secondary component about
3 arcsec north, suggests that either there is a counter jet or the
VLBI jet bends back to the north. 
\begin{figure}
\psfig{figure=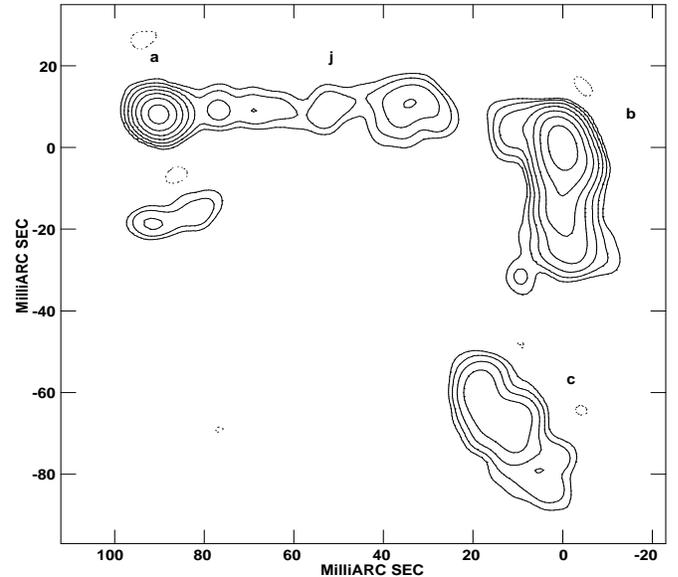,height=8.9cm,width=8.9cm}
\caption[]{ VLBI image of 0548$+$165 at 6\,cm. Contours are 
-1, 1, 2, 4, 6, 8, 16, 32, 64\,mJy/beam. The peak
flux density is 110.9\,mJy/beam.}
\end{figure}

{\bf 3C287} (1328+254)

\noindent
The image of 3C287 at $\lambda$6cm with 7 mas resolution shows a
regularly curving jet-like structure; this bears some similarity to
that in 3C119 (Fanti et al.\ 1989). The source brightness decreases
smoothly along the curved jet.   It is not clear where the core is
located so we cannot add its parameters to Table~3.  Fanti et al.\
1989 suggest that the main component (A in their Fig.\,3) is the 
possible site for the core.  This is the most compact feature visible
at $\lambda$6cm but it does have a spectral index of $\sim$0.5 between
$\lambda$18cm and $\lambda$6cm (Nan Ren-dong et al.\ 1988).

{\bf 1442$+$101} (OQ172)

\noindent
This object is unresolved by MERLIN. The VLBI image shows a very
compact source $\sim 70$ mas in extent at $\lambda$18cm (Dallacasa et
al.\ 1995).  It has a core-jet structure with the core located in the
northern part of the radio emission.  The source shows a bend in the
jet major axis PA of $90^{\circ}$ at a separation of $\sim15$\,mas.
1442$+$101 has a redshift of 3.544 and a very high integrated RM of 
22400 rad m$^{-2}$ in the source's rest frame (Taylor et al.\ 1992).
Recent VLBA observations at 5\,GHz by Udompresert et al.\ (1997)
indicate that the RM is 40,000 rad m$^{-2}$ in the rest frame of the
quasar. At 10 mas from the nucleus the RM falls to less than 100 rad
m$^{-2}$. The very high RM is found near to the core; it is likely
that this is not associated with material which could influence the
bending of the jet. 

{\bf 1629$+$680}

\noindent
This source was observed at $\lambda$13cm and $\lambda$3.6cm  
(Dallacasa et al.\ 1997).  The X-band image shows a straight jet $\sim
30$\,mas long. 
The S-band image shows a mild bend in the jet which finally changes 
the direction of the axis by $90^{\circ}$ at $100$\,mas from the core.

{\bf 1741$+$279}

\noindent
The VLA map at 8.4 GHz of 1741$+$279 shows two bright components
roughly aligned E-W, a wiggling jet 5 arcsec long aligned N-S to the 
north of the two bright components and a region of weak emission 4
arcsec to the south-south-east. The two compact components have about 
5\% of the total polarized emission at 8.4~GHz (Mantovani et al.\ 1997). 
The magnetic field is parallel to the curved line joining the two
components, changing direction smoothly by an angle of $\sim
90^{\circ}$; this suggests that there is a bend in the emitting region.
The VLBI $\lambda18$\,cm and $\lambda6$\,cm maps, Figures~3 \& 4
respectively, show several knots along the axis between these two
components.  The eastern  component is the nucleus (it has an inverted
spectrum) and the western one shows an elongation north-south.  When
compared with the $\lambda3.6$\,cm VLA image, it is possible to imagine
that this elongation is the location of a sharp bend or cusp in the jet's
apparent path. 
\begin{figure}
\psfig{figure=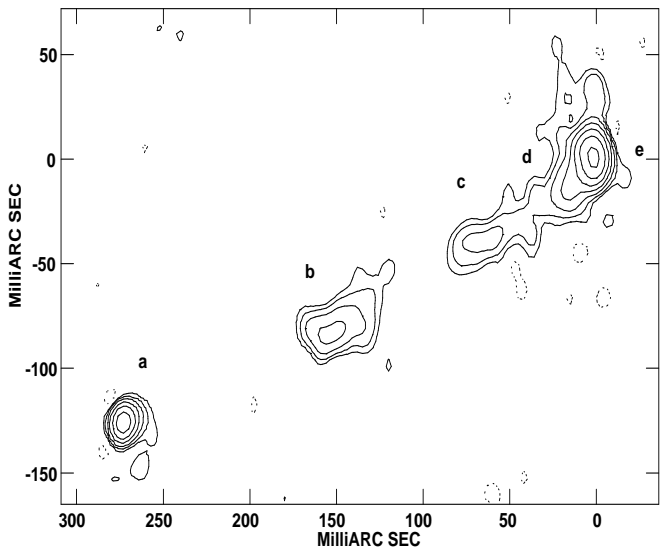,height=8.9cm,width=8.9cm}
\caption[]{ VLBI image of 1741$+$279 at 18\,cm. Contours are -1,
1, 2, 4, 8, 16, 32, 64\,mJy/beam. The peak
flux density is 77.3\,mJy/beam.}
\end{figure}
\begin{figure}
\psfig{figure=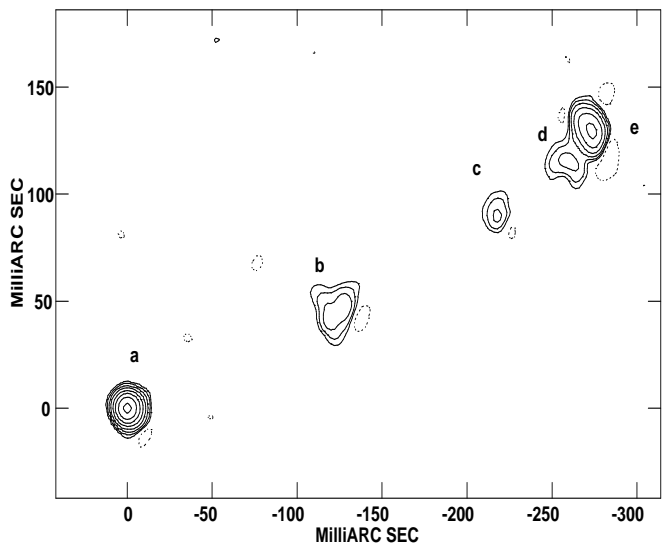,height=8.9cm,width=8.9cm}
\caption[]{ VLBI image of 1741$+$279 at 6\,cm. Contours are -0.5,
0.5, 1, 2, 4, 6, 8, 16, 32, 64\,mJy/beam. The peak
flux density is 75.9\,mJy/beam.}
\end{figure}

{\bf 2033$+$187}

\noindent
This source is unresolved by the VLA A-array at 15\,GHz (angular size
$< 0.05$\,arcsec) and is unpolarized.  We present two VLBI images here. The
first was obtained with a Global array at $\lambda18$\,cm (Fig.\,5); the
second with the EVN at $\lambda6$\,cm (Fig.\,6). In these images, we
see a straight jet and a dramatic bend at a small distance, $40$\,mas,
from component `a'.  Note that component `a' is the brightest feature
in 2033$+$187 at both frequencies and shows a steep spectral index
($\alpha=1.2$). The position of the core is unknown.  The resolution
in the north-south direction is insufficient to resolve the westermost 
component at $\lambda18$\,cm which has a low brightness region of
emission extending south. The $\lambda6$\,cm image shows a linear
morphology from component `a' to `d', where the jet changes direction
sharply. 
\begin{figure}
\psfig{figure=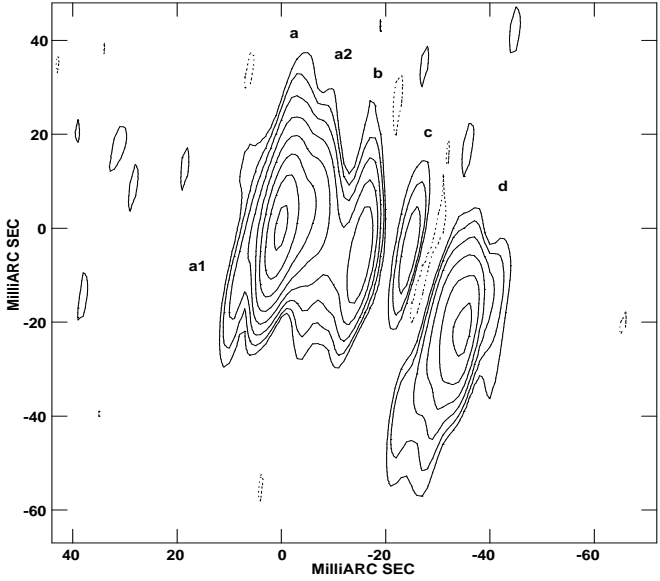,height=8.9cm,width=8.9cm}
\caption[]{ VLBI image of 2033$+$187 at 18\,cm. Contours are 
-2, 2, 4, 8, 16, 32, 64, 128, 256\,mJy/beam. The peak
flux density is 303.2\,mJy/beam.}
\end{figure}
\begin{figure}
\psfig{figure=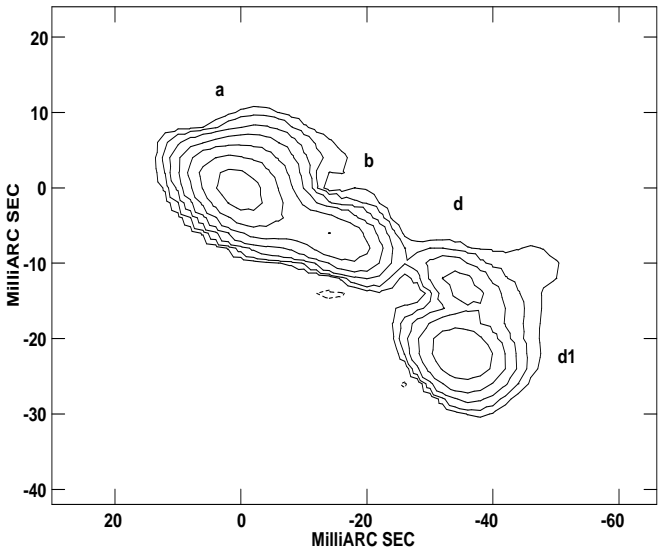,height=8.9cm,width=8.9cm}
\caption[]{ VLBI image of 2033$+$187 at 6\,cm. Contours are 
-1, 1, 2, 4, 8, 16, 32, 64\,mJy/beam. The peak
The peak flux density is 94.8\,mJy/beam.}
\end{figure}

{\bf 2147$+$145}

\noindent
VLBI observations of 2147+145 were made for the first time by Cotton 
et al.\ (1984) at $\lambda18$\,cm. That image shows a core-jet
structure with the major axis PA of 45$^{\circ}$.  Recent VLA images of  
2147+145 made at 8.4 and 15\,GHz, when compared with the previous
observations by Cotton (1983), show that the total flux density at
15~GHz has increased by $\sim$13\%.  In addition, a new component is
found north of the core in PA $-40^{\circ}$, separated by 0.35 arcsec
(Mantovani et al.\ 1997). This component lies in a direction which
differs by more than $80^{\circ}$ from that found for the jet in the
$\lambda18$\,cm VLBI map of Cotton {\it et al.} (1984).  The VLBI
source can be modelled by four Gaussian components that lie along a
path which bends smoothly towards north.  It is reasonable to expect
that the jet continues with an increasingly pronounced bend to allow
the flow to reach the component to the north. 

The EVN observations of 2147+145 at $\lambda$6\,cm (Fig.\,7) have
produced an image that shows a core-jet structure which, at first
sight, agrees with the $\lambda$18cm image of Cotton (1984).  A sharp
bend in the jet occurs at $\sim$40 mas from the core.  The flow
changes direction by $\sim90^{\circ}$ and a couple of weaker 
components are detected further north. The component `a', possibly the
core, has a steep spectrum.  The spectral index ranges from 0.38 (peak
emission) to 0.7 (total emission) between $\lambda$18cm and
$\lambda$6cm.   
\begin{figure}
\psfig{figure=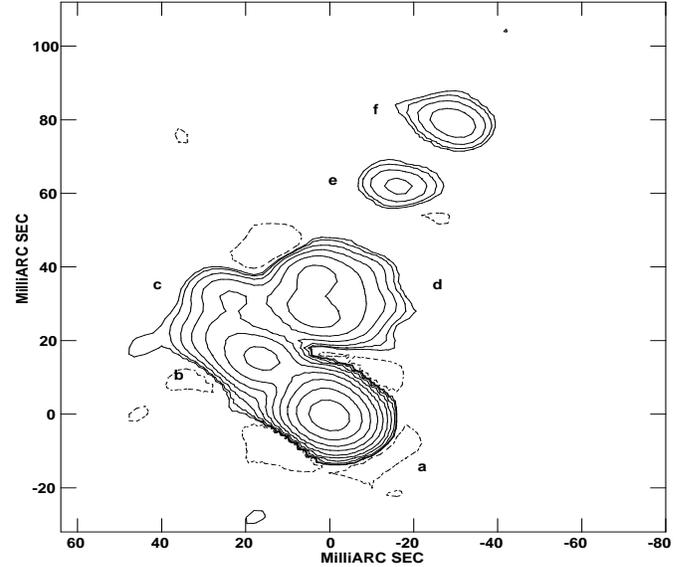,height=8.9cm,width=8.9cm}
\caption[]{ VLBI image of 2147$+$145 at 6\,cm. Contours are -0.6,
0.6, 1, 2, 4, 8, 16, 32, 64, 128, 256\,mJy/beam. The peak
flux density is 394.5\,mJy/beam.}
\end{figure}

\subsection{Summary of source parameters}

Some of the parameters for  the overall mas-scale structure that can be
derived from the available images of these sources are summarized in
Table~3. The table contents 
are as follows: -- column 1: name; column 2: references for the images; 
column 3: separation (in milliarcsecond)
between the core and the first large bend in the jet; column 4: linear
separation, assuming $H_{0}=100\,$km sec$^{-1}$Mpc$^{-1}$ and 
$q_{0} = 1$; column 5: difference
in the jet PA before and after the bend (in degrees); column 6: redshift;
column 7: optical identification; column 8: Rotation Measure in rad\,m$^{-2}$
corrected by the redshift. The symbol `--' means measurements are
unavailable.  The symbol `n' means polarization was not detected; 
; column 9: references for RMs.
%
%
%
\begin{table*} 
\small
\begin{center}
\caption{Source parameters.}
\medskip
\begin{tabular}{|l|c|c|r|c|c|c|r|c|} 
\hline  
Source&ref& Sep. & Sep. &$\Delta$PA& z & O.I. & RM$\times(1+z)^2$ & ref    \cr
      &   & mas  & pc   & deg      &   &      & rad m$^{-2}$ &             \cr
\hline 
0127$+$233 3C43   & a & 225 & 780 &  93 & 1.459 & Q & $-$1088     & 1 \cr
0358$+$004 3C99   & b &  15 &  46 &  60 & 0.425 & G &       0     & 2 \cr
0429$+$415 3C119  & c &  39 & 142 &  55 & 1.023 & Q &    3400     & 3 \cr
0538$+$498 3C147  & d & 200 & 664 &  90 & 0.545 & Q & $-$3144/630 & 4 \cr
0548$+$165        & e &  80 & 253 &  90 & 0.474 & Q &    1934     & 5 \cr
1328$+$254 3C287  & f &     &     &  58 & 1.055 & Q &    --       &   \cr
1442$+$101 OQ172  & g &  15 &  39 &  90 & 3.531 & Q &  22400      & 3 \cr
1629$+$680 4C68.18& h &  76 & 226 &  66 & 2.475 & Q &    --       &   \cr
1741$+$279 B2     & e & 300 & 862 &  84 & 0.372 & Q & $-$148/$-$279  & 2 \cr
2033$+$187        & e &  40 & 129 &  80 & (0.5) &-- &     n       & 2 \cr
2147$+$145        & e &  23 &  75 &  90 & (0.5) &-- &     n       & 2 \cr
\hline
\end{tabular}
\end{center}
%
{\em Note --}~~References - column 2 -: a. Spencer at al.\ (1991);
b. Mantovani et al.\ (1997) 
; c.Nan Ren-dong et al.\ (1991); d. Alef et al.\ (1990); e. this paper; f.
Fanti et al.\ (1989); g. Dallacasa et al.\ (1995); h. Dallacasa et al.\
(1997).  - column 9 -: 1. Junor et al.\
(in preparation); 2. Mantovani et al. (1997); 3. Taylor et al.\ (1992); 4. Junor 
et al.\ (1997); 5. Mantovani et al.\ (1995).
\end{table*}

Because of the selection criteria adopted, the sources listed in Table~3 
are characterized by the presence of bright jets. The change in the jet
major axis Position Angle has a mean value of $\Delta PA=78^\circ$.
The emission from the core compared to the total source emission is
weak ($<1\%$) in 3C43, 3C99, 3C119 and 3C147.   However, it is 
strong ($>5\%$) in 0548$+$165, 1629$+$680, 1741$+$279 and 2147$+$145. 
In two cases, namely 3C287 and 2033$+$187, it is still unclear where
the core is located.  It is worthwhile mentioning that all of the sources
in Table~3 have been identified with quasars --- with the exception of
3C99, which has been classified as an N-galaxy.  Note, too, that the
sources 2033$+$187 and 2147$+$145 are associated with unusually-faint
optical objects (Cotton et al.\ 1989).  

The sites where the bends occur are very close to the cores of the
sources. The angular separation from the core to that location
ranges between 15 and 300\,mas, which corresponds to a linear 
separation of 40--900\,pc, which is well inside the respective Narrow
Line Regions (NLRs).  

In general, most of the sources have a {\bf very} high Rotation
Measure. The sources are too compact to allow arcsecond-scale RM
images.  Those sources which are not polarized are also the most
compact among those listed; this might be interpreted as due to
large changes in the  magnetic field direction within the angular
resolution of the synthesized beam.  In such cases, VLBI polarimetry
is required; see, for example, the observations of 1442$+$101 
(Udomprasert et al.\ 1997)

\section{Statistical consideration}

The list of CSSs compiled by Dallacasa \& Stanghellini (1990) contains
$\sim$130 objects. About 30\% of these (i.e.\ $\sim39$ objects) are
dominated by the emission from the jet structure.  Table~3 contains
11 objects with strongly bent jets, which represents about 
30\% of this sub-sample of CSSs (or $\sim$8\% of the total number
of CSSs) which is a reasonably-significant number. 

If we consider instead the complete sample of CSSs defined by Fanti et
al.\ (1995) which contains 43 objects then five of them,
i.e.\ $\sim$11\%,  are listed in Table~3.  This confirms that a
strongly-bent jet is a noticeable phenomenon among this class of object.

As a comparison sample, we have taken the compilation of 293 Flat-Spectrum
Radio Sources constructed by Taylor et al.\ (1996) which gives a
complete Flux Density Limited VLBI sample.   Following Wilkinson
(1995), 82\% of objects in this sample are classified as asymmetric
core-jet sources.  We have found only 4 sources in this sample which
exhibit a bent jet fitting the selection criteria, and on the same
scales, that we have used for the CSSs listed in Table~3. 
This number represents less than 2\% of the objects in the complete
sample and is much less than that found for the CSS complete sample. 
This last number should be treated with some care, however. The
observations of the Flat Spectrum Sample were made with a high
resolution VLBI array, limited in sensitivity by the snap-shot 
observing strategy adopted.  Deeper observations might change the
number of objects with large bent jets. 

However, we consider the large fraction of objects with a strongly-bent jet
among CSSs as an indication that the propagation of the jet is
strongly affected by the ambient medium.

Let us calculate the probability of a collision between the jet and
the dense gas clouds in the NLR surrounding the radio source.
The outer gas is characterized by temperature of $\sim 10^{4}~K$, a
particle density of $\sim10^{4}$\,cm$^{-3}$ and by a filling factor
$\phi \leq 10^{-4}$ (McCarthy 1993).  The number of clouds,  {\rm B},
that the jet will encounter in a distance {\it L} is given by:
\begin{equation}
{\rm B} = d^2 \times L \times n = d^2 \times L \times {N\over L^3} =
{d^2\over L^2} \times N $$
\end{equation}
where {\it N} the total number of  clouds, {\it L} is the size of a
Narrow Line Region, {\it d} the diameter of a cloud, and {\it n} is
the cloud spatial density $N/L^3$.  The diameter {\it d} and the
density {\it n} of these dense gas clouds in active galaxies with
$z\geq0.1$ is still uncertain.  The term {\rm B} can be expressed more
appropriately in terms of the filling factor.

The radio jet propagates in a single preferred direction; thus we are
interested in determining the linear filling factor $\phi_l$, which is
defined as: 
\begin{equation}
\phi_l = {{\rm B} \times d \over L} = {d^2 \over L^2} \times N \times
{d \over L} = {d^3 \over L^3} \times N 
\end{equation}
The term on the right is, by definition, the {\it volume filling factor}
$\phi_v $. Such an expression is valid if the jet diameter is $< d$, 
the cloud's diameter. From this, one can derive $n = \phi_v/d^3$
and ${\rm B}$ can be calculated as:
\begin{equation}
{\rm B} = d^2 \times L \times {\phi_v \over d^3} = {L \phi_v \over d}$$
\end{equation}

To estimate {\rm B}, we have adopted the following values:  (a) the
linear size  {\it L} of the NLRs can be taken to be in the range
1--10~kpc.  This range is suggested by the bends which occur inside
the first kpc from the core and from the fact that we are dealing with
CSSs whose linear size cannot exceed 15--20 kpc by definition; 
(b) the filling factor $\phi_v$ is $10^{-4}$. However, from the 
luminosity in strong lines like H$\beta$ in the NLR ($10^{39}-10^{42}$
ergs\,s$^{-1}$ in Seyfert 2s) and electron density of 10$^3$ cm$^{-3}$,
the filling factor can reach values as large as 10$^{-2}$ (see, for
example, Peterson 1997); (c) In nearby AGN, the NLRs are often
partially resolved by the observations and the inferred sizes are
generally $\geq$100 pc. This gives an upper limit to the diameter $d$
for a cloud. In our own Galaxy, dense clouds are characterized by
sizes of 1--10 pc (Cowie \& Sangaila 1986; Cox \& Reynolds 1987). Thus
the cloud diameters can be taken to be in the range 10--100
pc. Moreover, from the VLBI observations we see that the jets are not
resolved in a direction transverse to the major axis, with an angular
resolution of $\sim 7~$mas (about 10 pc for $z=0.5$).

From these considerations, and assuming that the jet diameter is
$< d$, then B will fall in the range 0.1--0.01.  This last value
is not just the number of collisions between the radio jet 
and the dense gas cloud in a CSS.  It also represents the fraction of
sources which suffer for such an interaction.   Here we wish to stress
that the chances of such a collision are not very small. 

The Dallacasa \& Stanghellini (1990) list of CSSs contains more than one 
hundred objects.  The
percentage of CSSs which show large bends in the propagation direction
of   the radio jet is a number which is roughly in agreement with the
value derived here.   The above considerations also indicate that the
the probability of multiple bends is quite small.

\section{Discussion}

Sub-arcsecond resolution VLBI maps of CSSs often show that these
sources have strongly-distorted structures with recognizable jet-like
features (Fanti et al.\ 1986; Spencer et al.\ 1991), consistent with
strong dynamical interactions between the jets and the ambient media.  
Although the largest distortions are often seen in sources dominated by
jets, and there are suggestions that this might at least partly be due
to projection effects (e.g.\ Spencer 1994), there are clear indications 
that intrinsic distortions due to interactions with a possibly-dense
inhomogeneous gaseous environment play an important role. For example,
some of the most distorted and complex structures are found in objects
with very weak cores, as in the quasars 3C43 and 3C119. Such
objects should be inclined at large enough angles to the line of sight
that projection effects are not very significant.  Generally speaking,
CSSs usually show very weak cores compared to the jet emission while
in superluminal sources the Doppler boosted cores are the components
with the more dominant emission. Moreover,  the only CSSs known so
far which exhibit superluminal separation between two components,
thus implying a small angle of inclination to the line of sight, are 3C147
($v_{app}\sim1.3 c/h$; Alef et al.\ 1990) and 3C138 ($v_{app}\sim5
c/h$; Cotton et al.\ 1997).  In addition to the distorted structures,
an increased asymmetry in the location of the outer components also
suggests that the components are evolving through a dense asymmetric
environment in the central regions of these galaxies (Sanghera et
al.\ 1995; Saikia 1995). Such an asymmetry between the mas structure
and the arcsec structure is present where an arcsec structure is
detected for sources in Table~3. 

Most CSSs show low ($\sim$ 1$\%$) percentage polarizations at or below
5 GHz (Saikia 1991; Saikia, Singal \& Cornwell 1987). The median
polarization, however, increases with  frequency, (van Breugel et
al.\ 1992; Saikia 1995) suggesting that the polarimetric behaviour is
due to Faraday depolarization.  This pattern is consistent with the
observation of Garrington and Akujor (1996) that smaller sources show
stronger depolarization.  A number of sources with very high Rotation 
Measures in the range of several thousands of rad\,m$^{-2}$ are
also CSSs.  Most of the sources in Table~3 follow this trend having
large depolarization and do have rather large RMs. 

Sub-arcsec polarimetry has often provided evidence in favour of
the interaction of these components with dense clouds of gas. For
example, the southern component of 3C147 has a much higher RM than the
northern component, is brighter, closer to the nucleus, and has the
expected signatures of a jet colliding with a cloud of gas on the
southern side of the galaxy (Junor et al.\ 1997).   To explain both
the low polarizations at centimeter wavelengths and the small linear
sizes, it has been proposed that CSSs are cocooned in dense gaseous
envelopes (e.g.\  Mantovani et al.\ 1994).

Existing ground-based optical observations show
that narrow-line luminosities are comparable for CSSs and large-sized
radio sources. However, broader narrow-line profiles found for CSSs
suggest that the interaction of the jet with the ambient medium produces
additional accelerations (Gelderman 1992; Morganti 1993; Gelderman \&
Whittle 1994). From the information available in the literature, it is 
not clear whether there is enough gas to support the so-called
frustration scenario for CSSs, in which they are confined to small
dimensions by dense interstellar media, thus  hindering
their evolution into classical FRII radio sources. 

The structures of the sources shown here, with such large bent jets in the
vicinity of the core, support the view of a strong interaction between
the jet flow and dense gas clouds.  This is in agreement with our
estimate of the probability of having an interaction between the jet
and a dense cloud of gas inside the first kpc from the nucleus.  We
predict that the widths of the narrow emission lines in these sources
will be broader that for NLRGs (Narrow Line Radio Galaxies).  

The interaction between the radio source and the ISM has been investigated,
with time-dependent numerical simulations of jets associated with CSSs, 
by de Young (1993) in the following scenarios: {\it a)} the ISM is an uniform
homogeneous medium, with a smooth density gradient; {\it b)} the ISM
is in two phases --- a hot tenuous gas together with a population 
of cold dense clouds; {\it c)} the interaction is with a single large 
feature such a very dense cloud. de Young shows that CSSs of low and
intermediate luminosity can be confined by an ISM of average density
10 times that of the Galaxy. Confinement of the highest luminosity
sources requires an ISM of unrealistic density and total mass. The
population as a whole can be confined if the jets are intrinsically
more efficient, by factors of 10 or more, than is commonly assumed for
`normal' radio sources. 

The physics of jet-cloud collisions have been investigated by de Young
(1991) and Norman \& Balsara (1993) in 3-D hydrodynamical simulations.
Norman \& Balsara do identify a fluid dynamical mechanism which
maintains the collimation of the jet for all deflection angles
$0^{\circ} \leq \theta \leq 90^{\circ}$.  The reflected jet inherits
the stability properties of the original jet.  de Young shows that
deflection of a jet in the case of jet-loud encounters over the short
times ($\sim 10^6$ years) and distances (2 to 4 kpc) needed for CSSs
may easily occur. 

\section{Conclusions}

Fanti et al.\ (1995) have distinguished between double-lobed CSSs, which
represent $\sim70\%$ of the population and complex morphologied or asymmetric
jet-dominated CSSs. They suggest that double-lobed CSSs represent the
precursors of the large, double-lobed radio sources. The remaining
30\% are sources with the jets brightened by Doppler boosting or by
interaction with the ambient medium. The sources in Table~3 are classified
as core-jet objects. The evidence that Doppler boosting
can play a role in magnifying the large distortions seen in the jets is
very weak. Radio structures, large RMs, and broader narrow line
profiles support 
the view that large bent jets can be caused by interactions between the
jets and dense gas clouds in the NLRs. Even a rough statistical
computation indicates that the probability of a bump in the NLRs is 
statistically significant.  3-D hydrodynamical simulations of
extragalactic jets reveal interesting features that further support
this interpretation.  In this picture, sources like 3C119 and 3C287
require more than one collision to createstructures that are
spiral-like.  Such cases cannot be excluded statistically, though they
are very unlikely. 

\acknowledgements
{
The authors wish to thank the referee Dr.\ D.S.\ de Young for valuable
comments on the manuscript and the station and correlator staffs for
their collaboration for the data acquisition and correlation. The
National Radio Astronomy Observatory is operated by Associated
Universities Inc., under cooperative agreement with the National
Science Foundation. {\em AIPS} is NRAO's {\it Astronomical Image
Processing System}. {\em DIFMAP} was written by Martin Shepherd at
Caltech, and is part of the Caltech VLBI Software Package.
}
%
%
%
%

%
\end{document}